# Dynamics of Suspended Nanoparticles in a Time-varying Gradient Magnetic Field: Analytical Results

S.I. Denisov*, T.V. Lyutyy, A.T. Liutyi

*Sumy State University, Rymsky-Korsakov St., 2, 40007 Sumy, Ukraine*

We study theoretically the deterministic dynamics of single-domain ferromagnetic nanoparticles in dilute ferrofluids, which is induced by a time-varying gradient magnetic field. Using the force and torque balance equations, we derive a set of the first-order differential equations describing the translational and rotational motions of such particles characterized by small Reynolds numbers. Since the gradient magnetic field generates both the translations and rotations of particles, these motions are coupled. Based on the derived set of equations, we demonstrate this fact explicitly by expressing the particle position through the particle orientation angle, and vice versa. The obtained expressions are used to show that the solution of the basic set of equations is periodic in time and to determine the intervals, where the particle coordinate and orientation angle oscillate. In addition, this set of equations is solved approximately for the case of small characteristic frequency of the particle oscillations. With this condition, we find that all particles perform small translational oscillations near their initial positions. In contrast, the orientation angle oscillates near the initial angle only if particles are located in the vicinity of zero point of the gradient magnetic field. The possible applications of the obtained results in biomedicine and separation processes are also discussed.

**Keywords:** Dilute ferrofluids, Single-domain nanoparticles, Gradient magnetic field, Balance equations, Translational and rotational dynamics.

## 1. INTRODUCTION

The suspended single-domain ferromagnetic nanoparticles have interesting physical properties and potential biomedical applications such as cancer hyperthermia therapy, magnetic resonance imaging, targeted drug delivery and cell separation (for a recent review see, e.g., Refs. [1-4]). These applications are based on specific physical properties, which are mainly caused by the magnetic and mechanical (both translational and rotational) dynamics of such nanoparticles.

One of the most general approaches to the theoretical description of the coupled magnetic and mechanical dynamics uses the concept of the total angular momentum of nanoparticles. In particular, it has been used to study some features of the coupled magnetic and rotational dynamics arising from the interaction between the magnetic and lattice subsystems [5-8]. If the anisotropy magnetic field is large compared to the external fields, then the nanoparticle magnetization can be considered as "frozen" into the particle body [9]. In this case, the magnetization dynamics is completely determined by the nanoparticle rotation, and many of its characteristics in a circularly polarized magnetic field can be calculated analytically [10, 11].

The model of nanoparticles with "frozen" magnetization is also useful for studying the transport properties of suspended nanoparticles. Within this framework, we have predicted the phenomenon of their directed transport induced by the Magnus force in both deterministic and stochastic approximations (see Ref. [12] and references therein). Since the direction of motion and average velocity of nanoparticles can easily be controlled by external magnetic fields, the Magnus mechanism of directed transport could be used in drug delivery and separation applications. The model of "frozen" magnetization also allowed us to determine the transport properties of suspended nanoparticles subjected to a time-independent gradient magnetic field [13], which is often used in separation processes [14]. In the given work, we study analytically the coupled translational and rotational dynamics of such nanoparticles in a time-varying gradient magnetic field.

## 2. MOTION EQUATIONS

The translational and rotational motions of a suspended ferromagnetic nanoparticle, which are induced by a time-varying gradient magnetic field, depend on many factors. In particular, they depend on the particle magnetic moment, interaction between the lattice and magnetic subsystems, particle size, surface structure, liquid properties and gradient field characteristics, to name only a few. Therefore, in order to make the theoretical analysis tractable, we restrict ourselves to considering the simplest model for the nanoparticle dynamics.

### 2.1 Coupled Balance Equations

Our model is intended to describe the nanoparticle dynamics in dilute ferrofluids, when the interparticle interactions are negligibly small. The ferromagnetic nanoparticles are considered to be spherical with the same radius $a$. On the one hand, it is assumed to be so small that the single-domain state is realized and, on the other hand, so large that the influence of thermal fluctuations on the nanoparticle dynamics is negligible (see, e.g., Ref. [15]). If the particle material is magnetically uniaxial and the corresponding anisotropy field is strong enough, then the particle magnetization $\mathbf{M} = \mathbf{M}(t)$ ($|\mathbf{M}| = M$ = const) can be considered as "frozen" into the material [9]. Therefore, denoting by $\boldsymbol{\omega} = \boldsymbol{\omega}(t)$ the particle angular velocity, the dynamics of $\mathbf{M}$ can be described by the kinematic equation

$$\frac{d}{dt}\mathbf{M} = \boldsymbol{\omega} \times \mathbf{M}, \qquad (1)$$

where the sign $\times$ denotes the vector (cross) product.

---

* denisov@sumdu.edu.ua

Let us assume that in the Cartesian coordinate system $xyz$ characterized by the unit vectors $\mathbf{e}_x, \mathbf{e}_y, \mathbf{e}_z$ the time-varying gradient magnetic field $\mathbf{H}_g = \mathbf{H}_g(t)$ has only the $x$ component. Then $\mathbf{H}_g$ in the nanoparticle center $\mathbf{R} = \mathbf{R}(t)$ is given by

$$\mathbf{H}_g = \mathbf{e}_x g R_x \sin(\Omega t + \phi). \qquad (2)$$

Here, $g (> 0)$ is the magnetic field gradient, $R_x = R_x(t)$ is the $x$ component of $\mathbf{R}$, and $\Omega$ and $\phi \in (0, 2\pi)$ are the frequency and initial phase of $\mathbf{H}_g$, respectively. This field induces both the translational motion of the particle along the $x$ axis and its rotational motion. If the initial magnetization $\mathbf{M}(0)$ lies, e.g., in the $xy$ plane, then $\mathbf{M}(t)$ stays in this plane for all $t > 0$ (this result holds for the "frozen" magnetization):

$$\mathbf{M} = M(\mathbf{e}_x \cos\varphi + \mathbf{e}_y \sin\varphi), \qquad (3)$$

where $\varphi = \varphi(t)$ is the angle between the $x$ axis and the vector $\mathbf{M}$. From (1) and (3) it follows that $\boldsymbol{\omega} = \mathbf{e}_z \omega_z$ (the particle rotates about the $z$ axis) and $d\varphi/dt = \omega_z$.

In order to derive the equations describing the translational and rotational motions of the nanoparticle subjected to the time-varying gradient magnetic field (2), we neglect for simplicity the inertial effects and use the corresponding balance equations. Calculating for our case the driving force $\mathbf{f}_d = V(\mathbf{M} \cdot \partial/\partial\mathbf{R})\mathbf{H}_g$ [$V$ is the nanoparticle volume, the symbol $\cdot$ denotes the scalar (dot) product] and the driving torque $\mathbf{t}_d = V\mathbf{M} \times \mathbf{H}_g$, we reduce the force balance equation $\mathbf{f}_d + \mathbf{f}_f = 0$ to

$$\mathbf{e}_x MVg \cos\varphi \sin(\Omega t + \phi) + \mathbf{f}_f = 0 \qquad (4)$$

($\mathbf{f}_f$ is the friction force) and the torque balance equation $\mathbf{t}_d + \mathbf{t}_f = 0$ to

$$-\mathbf{e}_z MVgR_x \sin\varphi \sin(\Omega t + \phi) + \mathbf{t}_f = 0 \qquad (5)$$

($\mathbf{t}_f$ is the friction torque). If the particle dynamics is characterized by small translational and rotational Reynolds numbers, then (see, e.g., Ref. [16]) $\mathbf{f}_f = -\mathbf{e}_x 6\pi\eta a \, dR_x/dt$ ($\eta$ is the liquid dynamic viscosity) and $\mathbf{t}_f = -\mathbf{e}_z 8\pi\eta a^3 \omega_z$. Finally, taking into account that $\omega_z = d\varphi/dt$ and introducing the dimensionless time $\tau = \Omega t$, the dimensionless particle position $r_x = R_x/a$ and the dimensionless characteristic frequency of the particle oscillations

$$\nu = \frac{Mga}{6\eta\Omega}, \qquad (6)$$

from Eqs. (4) and (5) one obtains the following set of the coupled first-order differential equations:

$$\dot{r}_x = (4/3)\nu \cos\varphi \sin(\tau + \phi), \qquad (7)$$

$$\dot{\varphi} = -\nu r_x \sin\varphi \sin(\tau + \phi), \qquad (8)$$

where the overdot denotes derivative with respect to the dimensionless time $\tau$. The solution of these equations, the pair $\{r_x, \varphi\}$ of functions $r_x = r_x(\tau)$ and $\varphi = \varphi(\tau)$, is assumed to satisfy the initial conditions $r_{x0} = r_x(0)$ and $\varphi_0 = \varphi(0) \in (0, \pi)$.

## 3. THEORETICAL ANALYSIS

In spite of their apparent simplicity, Eqs. (7) and (8) are difficult to solve analytically. Therefore, here we present only some exact results following from this set of equations and solve it approximately for $\nu \ll 1$, $|r_{x0}| \lesssim 1$ and $\nu \ll 1$, $\nu|r_{x0}| \gtrsim 1$.

### 3.1 Exact Results

Let the pair $\{r_x, \varphi\}$ be the solution of Eqs. (7) and (8) for a predetermined set $(r_{x0}, \varphi_0, \phi, \nu)$ of the initial values and parameters, $r_{x0}$, $\varphi_0$, $\phi$ and $\nu$. Then, it can be straightforwardly checked that the pair $\{r_x, \pi - \varphi\}$ represents the solution of these equations for the set of parameters $(r_{x0}, \pi - \varphi_0, \pi + \phi, \nu)$. Similarly, one can verify that the solutions $\{-r_x, \pi - \varphi\}$ and $\{-r_x, \varphi\}$ correspond to the parameter sets $(-r_{x0}, \pi - \varphi_0, \phi, \nu)$ and $(-r_{x0}, \varphi_0, \pi + \phi, \nu)$, respectively. Thus, if, e.g., the dimensionless initial particle position $r_{x0}$ is changed to $-r_{x0}$, the initial phase $\phi$ is changed to $\pi + \phi$ and the other parameters $\varphi_0$ and $\nu$ are the same, then the solution of Eqs. (7) and (8) is given by the pair $\{-r_x, \varphi\}$.

The set of Eqs. (7) and (8) can also be represented in another form useful for the theoretical analysis of the particle dynamics. In order to derive this representation, we multiply Eq. (7) by $\sin\varphi$, Eq. (8) by $\cos\varphi$ and divide the second by the first. In doing so, one obtains

$$\frac{\dot{\varphi}\cos\varphi}{\sin\varphi} = -\frac{3}{4}r_x \dot{r}_x. \qquad (9)$$

Since by assumption $\varphi_0 \in (0, \pi)$, it is expected that the angle $\varphi = \varphi(\tau)$ also belongs to this interval. Using this condition and the relation $dr_x^2/d\tau = 2r_x \dot{r}_x$, Eq. (9) can be reduced to

$$\frac{d}{d\tau}\ln\sin\varphi = -\frac{3}{8}\frac{d}{d\tau}r_x^2. \qquad (10)$$

Finally, by integrating both sides of Eq. (10) with respect to $\tau$ from 0 to $\tau$, we get the equation

$$r_x^2 = r_{x0}^2 + \frac{8}{3}\ln\frac{\sin\varphi_0}{\sin\varphi} \qquad (11)$$

that connects $r_x$ and $\varphi$ and can be considered as the first equation for determining $\{r_x, \varphi\}$. Its important feature is that it is an algebraic, not differential, equation (it holds also for the time-independent gradient magnetic field, see Ref. [13]) The second equation, which is necessary to find $\{r_x, \varphi\}$, may be written, e.g., in the form of Eqs. (7), (8) or

$$\left(\frac{3\dot{r}_x}{4\sin(\tau+\phi)}\right)^2 + \left(\frac{\dot{\varphi}}{r_x \sin(\tau+\phi)}\right)^2 = \nu^2. \qquad (12)$$

Equation (11) permits us to make some general conclusions on the character of the nanoparticle dynamics. Indeed, since according to (11) $r_x = r_x(\varphi)$, the variables $\varphi$ and $\tau$ in Eq. (8) can be separated, yielding

$$\int_{\varphi_0}^{\varphi(\tau)} \frac{d\varphi}{r_x(\varphi)\sin\varphi} = -\nu F(\tau), \qquad (13)$$

where for convenience we introduced the notation

$$F(\tau) = \cos\phi - \cos(\tau + \phi). \qquad (14)$$

From this, it follows that the angle $\varphi$ is a periodic function of time: $\varphi(\tau) = \varphi(\tau + 2\pi n)$ $(n = 1,2, \ldots)$. Similarly, transforming Eq. (11) to

$$\sin\varphi = \sin\varphi_0 \exp[-3(r_x^2 - r_{x0}^2)/8], \qquad (15)$$

we make sure that $\varphi = \varphi(r_x)$ and, separating variables $r_x$ and $\tau$, Eq. (7) can be represented in the form

$$\int_{r_{x0}}^{r_x(\tau)} \frac{dr_x}{\cos\varphi(r_x)} = \frac{4}{3}\nu F(\tau), \qquad (16)$$

showing that $r_x(\tau) = r_x(\tau + 2\pi n)$. Although integrals in the left-hand sides of Eqs. (13) and (16) are not calculated analytically, these equations explicitly show that their solution, $\{r_x, \varphi\}$, is time-periodic. It is important to emphasize that the periodic solution is established immediately. In other words, the particle position $r_x$ and the magnetization angle $\varphi$ oscillate periodically (with the dimensionless period $2\pi$) for all $\tau \geq 0$.

This general result shows that the stationary solution of Eqs. (7) and (8), $\{0, \pi/2\}$, cannot be realized. The reason is that the particles with different $r_{x0}$ could reach the stationary state $r_x = 0$ only if they perform both the oscillatory and non-oscillatory motions. But, as shown above, only oscillatory motion is possible for particles in the time-varying gradient magnetic field. It should be noted that the impossibility of the stationary solution $\{0, \pi/2\}$ for arbitrary initial values $r_{x0}$ and $\varphi_0$ follows from Eq. (11) as well.

Using Eq. (11), it is also possible to estimate the angle interval, in which the magnetization angle oscillates at $|r_{x0}| \leq r_{cr}$, where the critical value of the initial particle position is defined as

$$r_{cr} = 2\left(\frac{2}{3}\ln\frac{1}{\sin\varphi_0}\right)^{1/2}. \qquad (17)$$

With this definition, the condition of non-negativity of the right-hand side of Eq. (11) can be written in the following form: $\sin\varphi \leq \exp[-3(r_{cr}^2 - r_{x0}^2)/8]$. Assuming that $\varphi_0 \in (0, \pi/2)$, from this condition one obtains $\varphi \in (0, \varphi_{cr})$, where the critical angle $\varphi_{cr}$ is given by

$$\varphi_{cr} = \arcsin \exp[-3(r_{cr}^2 - r_{x0}^2)/8] \qquad (18)$$

(note, $\varphi_{cr} \geq \varphi_0$ and $\varphi_{cr} = \varphi_0$ at $r_{x0} = 0$, $\varphi_{cr} = \pi/2$ at $|r_{x0}| = r_{cr}$). A similar consideration shows that if $\varphi_0 \in (\pi/2, \pi)$, then $\varphi \in (\pi - \varphi_{cr}, \pi)$.

An interesting feature of the critical angle $\varphi_{cr}$ is that it depends only on the initial particle position $r_{x0}$ and the initial magnetization angle $\varphi_0$, but not on the initial phase $\phi$ and the dimensionless characteristic frequency $\nu$ of the gradient magnetic field. As a consequence, the magnetization angle $\varphi$ oscillates within the same angle interval $(0, \varphi_{cr})$ [if $\varphi_0 \in (0, \pi/2)$] or $(\pi - \varphi_{cr}, \pi)$ [if $\varphi_0 \in (\pi/2, \pi)$] at arbitrary values of $\phi$ and $\nu$. This is a rather unexpected result because if, for example, $r_{x0} \in (0, r_{cr})$, $\varphi_0 \in (0, \pi/2)$ and $\phi = \pi$, then Eq. (8), at first glance, yields $\dot\varphi > 0$ for all $\tau \in (0, \pi)$. If so, then $\varphi$ may tend to $\pi$ as $\tau \to \pi$ if $\nu$ is large enough, i.e., $\varphi$ may exceed $\varphi_{cr}$. However, this reasoning is not correct because, according to Eq. (7), the particle position $r_x$ initially decreases with $\tau$, and if $r_x$ changes sign, then $\dot\varphi$ changes sign too. Thus, the restriction of the intervals, where the magnetization angle $\varphi$ oscillates at $|r_{x0}| \leq r_{cr}$, is a direct consequence of the complex coupled dynamics of $r_x$ and $\varphi$. Note also that for $|r_{x0}| > r_{cr}$ the above analysis becomes inapplicable (in this case the right-hand side of Eq. (11) is positive).

Unfortunately, we are not able to solve the set of Eqs. (7) and (8) exactly. Therefore, to illustrate some important features of the nanoparticle dynamics in the time-varying gradient magnetic field, below we solve this set of equations approximately.

### 3.2 Nanoparticle Dynamics at $\nu \ll 1$

If the characteristic frequency of the particle oscillations is small, i.e., $\nu \ll 1$, then, according to Eq. (7), all particles perform small translational oscillations near their initial positions. As to the rotational oscillations, their amplitude depends on the parameter $\nu r_x$ and thus on $\nu r_{x0}$, see Eq. (8). Therefore, we consider separately two cases, when the conditions 1) $|r_{x0}| \lesssim 1$ and 2) $\nu|r_{x0}| \gtrsim 1$ are held together with the condition $\nu \ll 1$.

*First case.* Let us represent the solution $\{r_x, \varphi\}$ of Eqs. (7) and (8) in the form $r_x = r_{x0} + r_{x1}$ and $\varphi = \varphi_0 + \varphi_1$, where the unknown functions $r_{x1} = r_{x1}(\tau)$ and $\varphi_1 = \varphi_1(\tau)$ satisfy the conditions $|r_{x1}|\sim\nu$, $|\varphi_1|\sim\nu$, $r_{x1}(0) = 0$ and $\varphi_1(0) = 0$. Then, keeping only the terms of the order $\nu$, these equations can be reduced to the set of uncoupled equations

$$\dot r_{x1} = (4/3)\nu \cos\varphi_0 \sin(\tau + \phi), \qquad (19)$$

$$\dot\varphi_1 = -\nu r_{x0} \sin\varphi_0 \sin(\tau + \phi). \qquad (20)$$

Solving Eqs. (19) and (20) with the initial conditions $r_{x1}(0) = 0$ and $\varphi_1(0) = 0$, we make sure that in the reference case, when $\nu \ll 1$ and $|r_{x0}| \lesssim 1$, the time dependence of the particle position is described as

$$r_x = r_{x0} + (4/3)\nu \cos\varphi_0 F(\tau) \qquad (21)$$

and the magnetization angle as

$$\varphi = \varphi_0 - \nu r_{x0} \sin\varphi_0 F(\tau). \qquad (22)$$

It can easily be verified that solutions (21) and (22) satisfy Eqs. (11) and (12) in the linear and quadratic approximations in $\nu$, respectively.

In accordance with the general result of the previous section, both variables $r_x$ and $\varphi$ oscillate with the period $2\pi$. As seen from (21) and (22), these oscillations occur about the average values

$$\langle r_x \rangle = r_{x0} + (4/3)\nu \cos\varphi_0 \cos\phi, \qquad (23)$$

$$\langle \varphi \rangle = \varphi_0 - \nu r_{x0} \sin\varphi_0 \cos\phi \qquad (24)$$

defined as $\langle (\cdot) \rangle = (1/2\pi)\int_0^{2\pi}(\cdot)d\tau$. Since by assumption $\nu \ll 1$, from (23) and (24) it follows that the average and initial values of $r_x$ and $\varphi$ differ only slightly.

Using (21), (22) and notation (14), we can also determine the maximum, $\max r_x$, and minimum, $\min r_x$,

values of the particle position $r_x$

$$\binom{\max}{\min} r_x = r_{x0} + \frac{4}{3}\nu \cos\varphi_0 \left[\cos\phi + \binom{+}{-}\mathrm{sgn}(\cos\varphi_0)\right] \quad (25)$$

and the maximum, $\max\varphi$, and minimum, $\min\varphi$, values of the magnetization angle $\varphi$

$$\binom{\max}{\min}\varphi = \varphi_0 - \nu r_{x0}\sin\varphi_0\left[\cos\phi - \binom{+}{-}\mathrm{sgn}(r_{x0})\right], \quad (26)$$

where the signum function $\mathrm{sgn}(\cdot)$ denotes the sign of its argument $(\cdot)$. We note in this regard that the angle intervals $(\min\varphi, \max\varphi)$, which correspond to $r_{x0} > 0$ and $r_{x0} < 0$, are determined by formula (26) much more precisely than those using the critical angle $\varphi_{\mathrm{cr}}$. The reason is that formula (18) (recall that it holds for $|r_{x0}| \leq r_{\mathrm{cr}}$) is determined for arbitrary values of the parameters $\phi$ and $\nu$, and so the angle intervals $(0, \varphi_{\mathrm{cr}})$ and $(\pi - \varphi_{\mathrm{cr}}, \pi)$ do not account for the specific features of the nanoparticle dynamics at $\nu \ll 1$ and $|r_{x0}| \lesssim 1$. Summarizing, we conclude that in this case rotations and translations of nanoparticles occur in small vicinities of $\varphi_0$ and $r_{x0}$.

*Second case.* If $\nu \ll 1$ and $\nu|r_{x0}| \gtrsim 1$ then, as before, we can represent the particle position as $r_x = r_{x0} + r_{x1}$, where, according to Eq. (7), $|r_{x1}| \sim \nu$. The last condition shows that the term $\nu r_x$ in Eq. (8) can be approximately replaced by $\nu r_{x0}$. Therefore, the nanoparticle dynamics in this case can be described by the set of simplified differential equations

$$\dot{r}_{x1} = (4/3)\nu\cos\varphi\sin(\tau + \phi), \quad (27)$$

$$\dot{\varphi} = -\nu r_{x0}\sin\varphi\sin(\tau + \phi). \quad (28)$$

In contrast to Eqs. (19) and (20), these equations are coupled and, since $\nu|r_{x0}| \gtrsim 1$, only Eq. (27) contains the small parameter $\nu$. In order to find the solution of Eqs. (27) and (28) with the initial conditions $r_{x1}(0) = 0$ and $\varphi(0) = \varphi_0$, we should first solve Eq. (28) and then, using the obtained solution, solve Eq. (27).

Equation (28) can readily be solved by the method of separation of variables. Indeed, using the table integral

$$\int \frac{dx}{\sin x} = \frac{1}{2}\ln\frac{1-\cos x}{1+\cos x},$$

from Eq. (28) one immediately obtains

$$\ln\frac{1-\cos\varphi}{1+\cos\varphi} - \ln\frac{1-\cos\varphi_0}{1+\cos\varphi_0} = -2\nu r_{x0}F(\tau). \quad (29)$$

Solving Eq. (29) with respect to $\cos\varphi$, after some simple algebra we arrive to the following expression:

$$\cos\varphi = \frac{\cos\varphi_0 + \tanh[\nu r_{x0}F(\tau)]}{1 + \cos\varphi_0\tanh[\nu r_{x0}F(\tau)]}. \quad (30)$$

It shows that, since $\nu|r_{x0}| \gtrsim 1$, the interval of oscillations of $\varphi$ is, in general, not small.

Now we substitute the right-hand side of (30) into Eq. (27). Then, integrating both sides of this equation over $\tau$ in the interval $(0, \tau)$ and taking into account that $r_{x1}(0) = 0$, we can write its solution in the form

$$r_{x1} = \frac{4\nu F(\tau)}{3\cos\varphi_0} - \frac{4\sin^2\varphi_0}{3r_{x0}\cos\varphi_0}\int_0^{\nu r_{x0}F(\tau)}\frac{dx}{1+\cos\varphi_0\tanh x}. \quad (31)$$

Although the definite integral in (31) can be calculated analytically for arbitrary values of $\varphi_0$ (see, e.g., Ref. [17], Eq. 2.448), the result is rather cumbersome. Therefore, here we analyze only three particular cases, when $\varphi_0 = 0, \pi/2$ and $\pi$. If $\varphi_0 = 0$ or $\pi$, then, according to (30) and (31), one gets

$$\cos\varphi|_{\varphi_0=0,\pi} = \pm 1, \quad r_{x1}|_{\varphi_0=0,\pi} = \pm(4/3)\nu F(\tau), \quad (32)$$

where the upper and lower signs correspond to $\varphi_0 = 0$ and $\varphi_0 = \pi$, respectively. These conditions show that nanoparticles do not rotate ($\varphi = 0$ at $\varphi_0 = 0$ and $\varphi = \pi$ at $\varphi_0 = \pi$) but only perform small translational oscillations in the vicinity of the initial position $r_{x0}$ ($r_x|_{\varphi_0=0,\pi} = r_{x0} + r_{x1}|_{\varphi_0=0,\pi}$). If $\varphi_0 = \pi/2$, then formulas (30) and (31) lead to the following conditions:

$$\cos\varphi|_{\varphi_0=\pi/2} = \tanh[\nu r_{x0}F(\tau)], \quad r_{x1}|_{\varphi_0=\pi/2} = 0 \quad (33)$$

(to avoid misunderstanding, we point out that the second condition in (33) is determined as the limit: $r_{x1}|_{\varphi_0=\pi/2} = \lim_{\varphi_0\to\pi/2} r_{x1}$). Thus, in contrast to the previous case, nanoparticles perform only the rotational oscillations, whose amplitude, as it follows from the first condition in (33), is not small (because $\nu|r_{x0}| \gtrsim 1$).

## 4. CONCLUSIONS

We have studied analytically both the translational and rotational dynamics of suspended ferromagnetic nanoparticles in a dilute ferrofluid subjected to the time-varying gradient magnetic field. Our approach to this problem is based on the main assumptions that (1) neglect inertial effects, (2) ignore thermal fluctuations, and (3) "freeze" the magnetization vector into each nanoparticle. Within these approximations, which hold for relatively large-sized nanoparticles with a strong uniaxial anisotropy, we have derived a coupled set of the first-order differential equations that describe the translational and rotational motions of such particles.

Using these equations, we have expressed the particle position, which describes the particle translational motion along the gradient magnetic field, through the magnetization angle, which describes the particle rotational motion. It has been explicitly demonstrated that this expression plays an important role in the theoretical analysis of the nanoparticle dynamics. In particular, using this expression, we have shown that the solution of the derived set of motion equations is always periodic (with the gradient field period) and estimated the intervals for the magnetization angle. In addition, to gain more insight into the nanoparticle dynamics, we have solved this set of equations for particles close to and far from the coordinate origin under the condition that the characteristic frequency of the particle oscillations is small. It turned out that while the translational oscillations are small for all particles, the rotational oscillations are small only for particles close to the origin.

Since the translational and rotational dynamics of ferromagnetic nanoparticles is responsible for many their properties, we expect that the theoretical results obtained in this paper will be useful for a number of modern applications. In particular, the resulting particle velocity, which is directly determined from the above results, can be used to calculate the power loss arising from the friction between nanoparticles and liquid. This contribution to the total power loss may be important, e.g., for magnetic hyperthermia applications. In addition, the theoretical analysis carried out here may be of interest also for drug delivery and separation applications. The reason is that the permanent magnetic field, which is applied perpendicular to the time-varying gradient magnetic field, could cause the directed transport of ferromagnetic nanoparticles along the gradient field.

## ACKNOWLEDGEMENTS

This work was partially supported by the Ministry of Education and Science of Ukraine under Grant No. 0119U100772.

## REFERENCES

1. *Nanoparticles for Biomedical Applications: Fundamental Concepts, Biological Interactions and Clinical Applications* (Eds. by E.J. Chung, L. Leon, C. Rinaldi) (Amsterdam: Elsevier: 2020).
2. *Nanoparticles and their Biomedical Applications* (Ed. by A.K. Shukla) (Singapore: Springer: 2020).
3. A. Farzin, S.A. Etesami, J. Quint, A. Memic, A. Tamayol, *Adv. Healthc. Mater.* 1901058 (2020).
4. S.D. Anderson, V.V. Gwenin, C.D. Gwenin, *Nanoscale Res. Lett.* **14**, 188 (2019).
5. K.D. Usadel, C. Usadel, *J. Appl. Phys.* **118**, 234303 (2015).
6. H. Keshtgar, S. Streib, A. Kamra, Ya.M. Blanter, G.E.W. Bauer, *Phys. Bev. B* **95**, 134447 (2017).
7. T.V. Lyutyy, O.M. Hryshko, A.A. Kovner, *J. Magn. Magn. Mater.* **446**, 87 (2018).
8. T.V. Lyutyy, S.I. Denisov, P. Hänggy, *Phys. Bev. B* **100**, 134403 (2019).
9. G. Bertotti, I.D. Mayergoyz, C. Serpico, *Nonlinear Magnetization Dynamics in Nanosystems* (London: Elsevier: 2009).
10. T.V. Lyutyy, S.I. Denisov, V.V. Reva, Yu.S. Bystrik, *Phys. Rev. E* **92**, 042312 (2015).
11. P. Ilg, A.E.A.S. Evangelopoulos, *Phys. Rev. E* **97**, 032610 (2018).
12. S.I. Denisov, T.V. Lyutyy, V.V. Reva, A.S. Yermolenko, *Phys. Rev. E* **97**, 032608 (2018).
13. S.I. Denisov, T.V. Lyutyy, M.O. Pavlyuk, *J. Phys. D: Appl. Phys.* **53**, 405001 (2020).
14. J. Svoboda, *Magnetic Techniques for the Treatment of Materials* (Dordrecht: Kluwer: 2004).
15. A.P. Guimarães, *Principles of Nanomagnetism, 2$^{nd}$ Edition* (Cham: Springer: 2017).
16. Z. Zapryanov, S. Tabakova, *Dynamics of Bubbles, Drops and Rigid Particles* (Dordrecht: Springer: 1999).
17. I.S. Gradshteyn, I.M. Ryzhik, *Table of Integrals, Series, and Products, 8$^{th}$ Edition* (Ed. by D. Zwillinger) (Oxford: Elsevier Academic Press: 2014).

## Динаміка зважених наночастинок у змінному в часі градієнтному магнітному полі: Аналітичні результати

С.І. Денисов, Т.В. Лютий, А.Т. Лютий

*Сумський державний університет, вул. Римського-Корсакова, 2, 40007 Суми, Україна*

Теоретично вивчається детерміністична динаміка однодоменних феромагнітних наночастинок у розбавлених ферорідинах, що знаходяться під впливом періодичного у часі градієнтного магнітного поля. Використовуючи рівняння балансу сил та моментів виведено систему двох диференційних рівнянь першого порядку, що описують трансляційний та обертальний рухи таких частинок у випадку малих чисел Рейнольдса. Оскільки градієнтне магнітне поле генерує як трансляційний, так і обертальний рухи частинок, ці рухи пов'язані між собою. Цей факт продемонстровано шляхом отримання за допомогою знайденої системи рівнянь співвідношень, що виражають положення частинки через кут її орієнтації, і навпаки. Отримані співвідношення використані, щоб показати, що розв'язок базової системи рівнянь є періодичним у часі, і щоб знайти інтервали, в яких відбуваються осциляції положення та кута орієнтації частинок. Крім цього, знайдено наближений розв'язок даної системи рівнянь у випадку, коли характерна частота коливань частинок мала. Встановлено, що в цьому випадку всі частинки здійснюють малі коливання поблизу початкових положень. В той же час, малі коливання кута орієнтації відносно початкового кута відбуваються лише для частинок, що знаходяться поблизу початку координат, де градієнтне магнітне поле мале. Обговорюється також можливе використання отриманих результатів в біомедицині та процесах сепарації наночастинок.

**Ключові слова:** Розбавлені ферорідини, Однодоменні наночастинки, Градієнтне магнітне поле, Рівняння балансу, Трансляційна та обертальна динаміка.